\documentclass[aps,pre,twocolumn,groupedaddress,amsmath,amssymb,nofootinbib]{revtex4-1}

\usepackage{graphicx,color}
\usepackage{caption}
\usepackage{subcaption}
\usepackage{amsmath}
\usepackage{epstopdf}
\usepackage{soul}

\newcommand{\D}{\mathrm{d}}
\newcommand{\e}{\mathrm{e}}

\newcommand{\be}{\begin{equation}}
\newcommand{\ee}{\end{equation}}
\newcommand{\bea}{\begin{eqnarray}}
\newcommand{\eea}{\end{eqnarray}}
\newcommand{\eps}{\varepsilon}

\newcommand{\kbt}{k_{\mathrm{B}}T}

\newcommand{\lb}{l_\mathrm{B}}
\newcommand{\ld}{\lambda_\mathrm{D}}
\newcommand{\lgc}{l_\mathrm{GC}}

\newcommand{\vecr}{{\bf r}}

\newcommand{\ra}[1]{\textcolor{black}{#1} } 

\newcommand{\ZO}{\zeta_{1}}
\newcommand{\ZT}{\zeta_{2}}
\newcommand{\halpha}{\frac{\alpha}{2}}

\begin{document}


\title{Electrostatic Attraction between Overall Neutral Surfaces}
\author{Ram M. Adar, David Andelman}
\email{andelman@post.tau.ac.il}
\affiliation{Raymond and Beverly Sackler School of Physics and Astronomy\\ Tel Aviv
University, Ramat Aviv, Tel Aviv 69978, Israel}
\author{Haim Diamant}
\affiliation{Raymond and Beverly Sackler School of Chemistry\\ Tel Aviv
University, Ramat Aviv, Tel Aviv 69978, Israel}

\date{July 25, 2016}

\begin{abstract}

Two overall neutral surfaces with positive and negative charged domains (``patches'') have been shown in recent experiments to exhibit long-range attraction when immersed in an ionic solution. Motivated by the experiments, we calculate analytically the osmotic pressure between such surfaces within the Poisson-Boltzmann framework, using a variational principle for the surface-averaged free energy. The electrostatic potential, calculated beyond the linear Debye-H\"uckel theory, yields an {\it overall attraction} at large inter-surface separations, over a wide range of the system's controlled length scales. In particular, the attraction is stronger and occurs at smaller separations for surface patches of larger size and charge density. In this large patch limit, we find that the attraction-repulsion crossover separation is inversely proportional to the square of the patch charge-density and to the Debye screening length.
\end{abstract}

\maketitle

\section{Introduction}
\label{sec1}
Long-range interactions between charged surfaces play an important
role in electrochemistry, materials science and biology~\cite{Israelachvily,VO,David95}.
For surfaces bounding an ionic solution, such interactions are
governed by the entropy and electrostatics of the ionic solutes and polar solvent. A standard tool to analyze the underlying physics
of these systems is the Poisson-Boltzmann (PB) theory~\cite{David95}. This is a mean field (MF) theory, within which ions are treated as point-like particles obeying a Boltzmann distribution, the aqueous solution
is taken as a continuous and homogeneous dielectric medium, and
\ra{in most treated cases}, the bounding surfaces are assumed to be homogeneously charged~\cite{Israelachvily,VO,David95,Ohshima,Dan07,SamEPL}.
However, as many charged surfaces in soft and biological matter are heterogeneous over
mesoscopic length scales, several experimental~\cite{Christension01,Perkin05,Meyer05,Zhang05,Perkin06,Meyer06,Hammer10,SilbertPRL,Popa10,Drelich11} and theoretical~\cite{Richmond74,Richmond75,Muller83,Kostoglou92,Milkavcic94,Holt97,Khachatourian98,Stankovitch99,Velegol01,Milkavcic95,White02,Lukatsky02,Lukatsky02b,Fleck05,Landy10,Naydenov07,Jho11,Velichko06,Rudi05,Rudi06,Rudi08,DanPRE,JoePPM,LevinMC,Maduar13} studies have investigated the effects of surface-charge heterogeneity on the inter-surface electrostatic
interactions.

\ra{For two surfaces with identical non-zero net charge and a small charge modulation, it was shown that the modulation has little effect as compared to that of the net charge \cite{Milkavcic95,White02,Lukatsky02,Lukatsky02b,Fleck05,Landy10}. For two {\it overall  neutral} surfaces, on the other hand, the effect of charge modulation is expected to be substantial~\cite{Richmond74,Richmond75,Muller83,Kostoglou92,Milkavcic94,Holt97,Khachatourian98,Stankovitch99,Velegol01,Naydenov07,Jho11,Velichko06,Rudi05,Rudi06,Rudi08,DanPRE,JoePPM,LevinMC,Maduar13}.}
Several experimental studies have examined the interaction between two such overall neutral surfaces, made of positive and negative domains (``patches'') that are much larger than the molecular size~\cite{Christension01,Perkin05,Meyer05,Zhang05,Perkin06,Meyer06,Hammer10,SilbertPRL}. \ra{For example, negative mica surfaces can be coated with neutralizing cationic surfactants that later dissociate and form positively charged bilayer patches.} An unexpected attraction was measured between two such surfaces~\cite{Christension01,Perkin05,Meyer05,Zhang05,Perkin06,Meyer06,Hammer10,SilbertPRL}.   \ra{At large inter-surface separations (beyond tens of nanometers), hydrophobic and dispersion interactions were ruled out as a possible origin for this attraction. Rather, it was shown to stem from the electrostatic interactions between the surface charges \cite{Meyer06,Hammer10}.}

\ra{ For systems with relaxation times shorter than the measurement time scale, the electrostatic attraction is explained by the self-adjusting of surface charges. Then, positively charged patches on one surface position themselves against negatively charged ones on the second surface and vice versa~\cite{Naydenov07,Jho11,Velichko06}. This is referred to as the {\it annealed} case. However, sometimes the surface charges are effectively "frozen" in time and the patch arrangements are random.  This is referred to as the {\it quenched} case. This scenario was tested in Ref.~\cite{SilbertPRL} by applying a relative shear velocity between the two surfaces, and observing that the patches do not rearrange on experimental time scales.}

Remarkably, the attraction effect prevailed even in the quenched case~\cite{SilbertPRL}. Unaware of the experimental results, one might have predicted the exact opposite: on average, the electrostatic effects for neutral surfaces are expected to cancel out, leaving a predominant entropic repulsion due to mobile ions. This was indeed found theoretically~\cite{DanPRE}, while employing the linear Debye-H\"uckel (DH) limit of the PB theory for any patch size and random arrangement. Only repulsion was obtained and the theory failed to capture the attraction effect.

Beyond the DH treatment, we single out three recent theoretical works. In the first, fluctuation effects were incorporated in a loop expansion of the free energy, going beyond the PB theory, for `molecular-size patches'. However, only a repulsive interaction was found~\cite{Rudi06} between the two bounding surfaces. For infinitely large patches, on the other hand, attraction was predicted in the work of Silbert {\it et al.}~\cite{SilbertPRL}. In their model, the system is averaged over two situations of two infinite and homogeneously charged surfaces facing each other. In the first, the surfaces are equally charged, while in the second one, the surfaces are oppositely charged. Their numerical calculation has shown that the repulsion in the former case is weaker than the attraction in the latter, yielding an overall attraction for the average between the two. This is due to the fact that counterions between oppositely charged surfaces are released into the bulk more freely, enhancing the electrostatic attraction~\cite{SamEPL}. However, this heuristic model does not retain the dependence of the osmotic pressure on the finite patch size that has great experimental relevance.

In a more recent work, Monte Carlo (MC) simulations were carried out for finite size surfaces that were divided into two or four homogeneously charged patches~\cite{LevinMC}. An attraction was found in both cases and was stronger for the larger patches. Since the patch size is still comparable with the finite system size, this numerical study offers only a limited insight into the patch-size dependence of the osmotic pressure. Consequently, no general relations with other relevant length scales were derived.

To account analytically for the long-range attraction between overall-neutral quenched patchy surfaces, we introduce in this paper a theoretical framework that substantially improves on the qualitative trends just described. In particular, our theory addresses two  previously-unanswered key questions: (a) what are the conditions for the existence of such inter-surface attraction? (b) How strong can this attraction be? Defining a parameter that combines the patch charge-density, ionic strength and patch size, enables us to derive  closed-form expressions for the osmotic pressure between two patchy surfaces in the limit of small and large inter-surface separations. Subsequently, we obtain the separation at which the interaction crosses over from repulsive to attractive, and its dependence on other system parameters. The existence of attraction is demonstrated for a wide range of parameters, and we conclude that it is the rule rather than the exception.

The outline of our paper is as follows. The model is
formulated in Sec.~\ref{sec2} alongside the analytic framework for the calculations. The results for the osmotic pressure between patchy surfaces are presented in Sec.~\ref{sec3}. In particular, the conditions for attraction are derived and the magnitude of the attraction is compared to the competing van der Waals attractive force. Finally, in Sec.~\ref{sec4} we discuss the implications of our results.

\section{Model}
\label{sec2}
\begin{figure}[h]
\centering
\includegraphics[width=0.85\columnwidth]{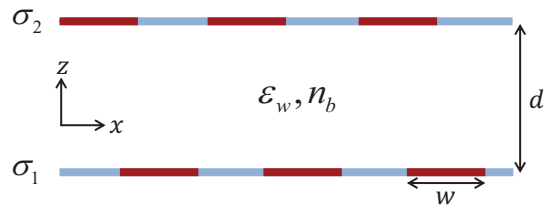}
\caption{(Color online) Schematic drawing of two patchy planar surfaces bounding an ionic solution with dielectric constant $\varepsilon_{w}$. Grey regions are positively charged and red (dark) regions are negatively charged. The surface-charge density is approximated  by a single $k$-mode modulation, Eq.~(\ref{eq1}), along the $x$-direction. The system is coupled to a reservoir of monovalent salt ions of concentration $n_{b}$.}
\label{fig1}
\end{figure}

Consider an aqueous solution confined between two planar
surfaces whose area, $A$, is taken to be arbitrarily large. The surfaces are located at $z=\pm d/2$,
where the $z$-axis is normal to the surfaces (see Fig.~\ref{fig1}). The solvent (water) is modeled as a homogeneous medium with a dielectric constant $\varepsilon_{w}$, and
is coupled to a reservoir of monovalent salt ions of concentration $n_{b}$. The medium
outside the surfaces is assumed to have a much lower dielectric constant and does not contain any ions.
Therefore, the electric field is confined to the aqueous solution bounded between the two surfaces.
The patchy surface-charge density is modeled by alternating
positive and negative stripes in the $x$-direction, and is assumed to be quenched. The stripes on the
bottom and top surfaces have an identical width, $w$, but are not commensurate. We approximate the surface charge profiles by a single $k$-mode modulation,
\begin{align}
\label{eq1}
\sigma_{1}(x) & =\sigma\cos(kx-\alpha/2)\nonumber \\
\sigma_{2}(x) & =\sigma\cos\left(kx+\alpha/2\right),
\end{align}
where $\sigma_{1}$ ($\sigma_{2}$) are the quenched bottom (top) surface charge densities, $\sigma$ is the patch charge-density, $k=2\pi/(2w)$ is the modulation wavenumber, and $\alpha$ is the relative phase between the two surface modulations. \ra{By the choice of Eq.~(\ref{eq1}) for the surface-charge densities, we limit ourselves to discuss only overall neutral surfaces with a quenched patch arrangement, as found experimentally, e.g., in Ref.~\cite{SilbertPRL}. Possible generalizations of Eq.~(\ref{eq1}) to other surface-charge densities will be discussed below (see also Appendix \ref{appB})}.

The \ra{osmotic pressure} between the surfaces changes with $\alpha$. In particular,
fully in-phase surfaces $\left(\alpha=0\right)$ repel, and out-of-phase ones $\left(\alpha=\pi\right)$  attract, demonstrating the major effect of the correlation between the top and bottom surfaces. Since we want to treat randomly charged surfaces, any correlations
between the two surfaces are removed by averaging over $\alpha$. For surfaces that are prepared independently with no inter-correlation, $\alpha$ gets any value, $0\le\alpha<2\pi$, with the same probability. The results can be generalized to any distribution of the relative phase, $\alpha$.

The free energy of the system can be derived in two equivalent methods. First, one can use the {\it charging method}~\cite{VO} and equate the free energy with the work required to increase the surface charge incrementally, at each point on the surface, from zero to the desired final value. For the surface charge densities of Eq.~(\ref{eq1}), we obtain
\begin{align}
  \label{eq3}
  F & =\int \D x\,\D y\,\int_{0}^{1}\psi\left(-d/2;\ZO\sigma,0\right)\sigma\cos\left(kx-\alpha/2\right)\D\ZO\nonumber \\
 & +\int \D x\,\D y\,\int_{0}^{1}\psi\left(d/2;\sigma,\ZT\sigma\right)\sigma\cos\left(kx+\alpha/2\right)\D\ZT,
\end{align}
where $\psi(\pm d/2;\sigma,\sigma')$ is the electrostatic potential at the $z=\pm d/2$ surfaces, given that the bottom (top) surface-charge density amplitude is $\sigma$ ($\sigma'$). The parameters $0\le \zeta_{1,2} \le 1$ describe the charging state of the bottom and top surfaces, respectively. They vary from $\zeta=0$ for an uncharged surface to $\zeta=1$ for a fully charged one.

Equivalently, one can derive~\cite{David95} the excess bulk free energy over that of a homogeneous electrolyte reservoir of concentration $n_{b}$ and with $\psi=0$.
 Using the thermodynamic relation $F=U-TS$ and inserting the electrostatic energy for $U$ and the ion entropy of mixing for $S$, one obtains the following form (employing Gaussian units):

 \begin{align}
 \label{eq2a}
 F&=\int \D ^{3}r \bigg[-\frac{\eps_{w}}{8\pi}\left(\nabla\psi\right)^{2}+\left(n_{+}-n_{-}\right)e\psi\nonumber\\
 &+\bigl[\sigma_{1}\delta(z+d/2)+\sigma_{2}\delta(z-d/2)\bigr]\psi \nonumber\\
 &+\kbt\sum_{\alpha=\pm}^{}\left(n_{\alpha}\ln\left(\frac{n_{\alpha}}{n_{b}}\right)-\left(n_{\alpha}-n_{b}\right)\right)\bigg],
 \end{align}
where $n_{\pm}$ are the concentrations of the positive and negative ions, $\delta(z)$ is the Dirac delta function and $\kbt$ is the thermal energy. Minimizing Eq.~(\ref{eq2a}) with respect to the fields $n_{\pm}$ yields the Boltzmann distribution, $n_{\pm}(\vecr)=n_{b}\exp(\mp\Psi(\vecr))$, where $\Psi=e\psi/\kbt$ is the dimensionless electrostatic potential. Inserting the Boltzmann distribution in Eq.~(\ref{eq2a}) and considering a charging state described by the parameters $\zeta_{1,2}$, we find the dimensionless free energy
\begin{align}
\label{eq2}
\frac{F}{\kbt} &=\left(8\pi\lb\right)^{-1}\int\D^{3} r \Big[-\left(\nabla\Psi\right)^{2}+2\ld^{-2}\left(1-\cosh(\Psi)\right) \nonumber \\
&+\frac{4}{\lgc }\zeta_{1}\cos\left(kx-\alpha/2\right)\delta\left(z+d/2\right)\Psi \nonumber \\
&+\frac{4}{\lgc }\zeta_{2}\cos\left(kx+\alpha/2\right)\delta\left(z-d/2\right)\Psi\Big],
\end{align}
where $\lb=e^{2}/\left(\varepsilon_{w} \kbt\right)$ is the Bjerrum length, $\ld=1/\sqrt{8\pi n_{b}\lb}$ is the Debye screening length, and $\lgc =e/\left(2\pi|\sigma| \lb\right)$
is the Gouy-Chapman length.

Equations (\ref{eq3}) and (\ref{eq2})  are equivalent~\cite{David95} and both are useful in our derivation. Namely, Eq.~(\ref{eq2}) is written in terms of a free energy density, and $\Psi$ is found from its variation. On the other hand, once $\Psi$ is found it is more convenient to use Eq.~(\ref{eq3}) for the calculation of the free energy, as it involves only the values of $\Psi$ at the surfaces. Furthermore, from Eq.~(\ref{eq3}) we see that if one decomposes $\Psi$ in a Fourier expansion in the $x$ direction, only the same $k$-mode of the charge modulation, Eq.~(\ref{eq1}), contributes to the free energy. Therefore, searching for a function $\Psi$ that minimizes Eq.~(\ref{eq2}), we consider only functions of the form: $\cos(kx)h(z)+\sin(kx)g(z)$. Substituting this variational ansatz in Eq.~(\ref{eq2}), we would like to integrate over the $x$-coordinate in order to obtain a free energy density in the $z$-coordinate. Then, the functions $h$ and $g$ that minimize the free energy can be found by solving the Euler-Lagrange equations. The problem, however, is that the integration over the $x$ coordinate cannot be done analytically. We overcome this obstacle by expanding $h$ and $g$ in a perturbative expansion in powers of a dimensionless parameter of the system.

For homogeneously charged surfaces, the system is well described by two dimensionless ratios, $d/\ld$ and $\ld/\lgc$. The ratio $d/\ld$ characterizes the inter-surface separation, while $\ld/\lgc$ characterizes the strength of the interaction that is enhanced in the presence of surface charge and diminished in the presence of the salt. When $\ld/\lgc$  is sufficiently small, the DH theory is applicable, yielding a linear dependence of the potential on this ratio. Equivalently, the DH theory is obtained as the 1st order of the expansion in $\ld/\lgc$.

For a single $k$-mode modulation of the surface charge, the system is described in a similar manner, replacing $\ld$ by a modified screening length that incorporates the surface charge modulation \ra{(for example, see  Refs. \cite{Ohshima} and \cite{Richmond74}).}
 \begin{equation}
 \label{eq4}
 p=\frac{1}{\sqrt{k^{2}+\ld^{-2}}}=\frac{w\ld}{\sqrt{\pi^2\ld^2+w^2}}<\ld.
 \end{equation}
The role of $d/p$ and $p/\lgc$ in the DH limit is the same as in the homogeneous case. Namely, $p/\lgc$ characterizes the strength of the interaction and the DH theory is obtained as the 1st order of the expansion in $p/\lgc$. As we search for corrections to the DH treatment that can give rise to attraction, we use $p/\lgc$ as our perturbative parameter and go beyond the 1st order in the expansion. Also, note that the additional length scale, $k^{-1}$, yields a third ratio, $p/\ld$, that characterizes the role of the patch size. It vanishes in the limit of infinitesimal patches ($p\to0$), and approaches unity in the opposite limit of infinite patches ($p\to\ld$).

In light of the aforementioned arguments, we substitute $\Psi$ in Eq.~(\ref{eq2}) using the following variational Ansatz
\begin{align}
\label{eq5}
\Psi\left(x,z\right) & =\frac{p}{\lgc }\left[\cos(kx)h\left(z/p\right)+\sin(kx)g\left(z/p\right)\right].
\end{align}
We expand the integrand of Eq.~(\ref{eq2}) up to the $4^{\rm{th}}$ order in $p/\lgc$ and perform the integration over $x$ and $y$, keeping only terms proportional to the surface area, $A$, which is taken to be arbitrarily large. The resulting Euler-Lagrange equations for $h$ and $g$ are
\begin{align}
\label{eq6a}
h''(\tilde{z})-h(\tilde{z})&=\frac{1}{8}
\left(\frac{p}{\lgc}\right)^{2}\left(h^{2}+g^{2}\right)h \nonumber\\
&-2\cos\left(\frac{\alpha}{2}\right)
\left[\zeta_{1}\delta\left(\tilde{z}+\frac{d}{2p}
\right)+\zeta_{2}\delta\left(\tilde{z}-\frac{d}{2p}\right)\right],\nonumber\\
g''(\tilde{z})-g(\tilde{z})&=\frac{1}{8}\left(\frac{p}{\lgc}\right)^{2}
\left(h^{2}+g^{2}\right)g \nonumber\\
&-2\sin\left(\frac{\alpha}{2}\right)
\left[\zeta_{1}\delta\left(\tilde{z}+\frac{d}{2p}\right)-
\zeta_{2}\delta\left(\tilde{z}-\frac{d}{2p}\right)\right],
\end{align}
where we rescale the $z$ axis by $p$, $\tilde{z}=z/p$. These coupled nonlinear equations can be transformed into decoupled linear equations by expanding $\Psi$ in orders of $p/\lgc$ according to
\begin{align}
\label{eq6}
\Psi\left(x,\tilde{z}\right) & =\cos\left(kx\right)\sum_{n\ \rm{odd}}\left(\frac{p}{\lgc }\right)^{n}h_{n}\left(\tilde{z}\right) \nonumber\\
&+\sin\left(kx\right)\sum_{n\ \rm{odd}}\left(\frac{p}{\lgc }\right)^{n}g_{n}\left(\tilde{z}\right),
\end{align}
where only odd powers are considered, since the potential is odd in the patch charge density, $\sigma\sim\lgc^{-1}$.
The 1st-order terms, $h_{1}$ and $g_{1}$, reproduce the DH solution, while the leading corrections in $\Psi$, of order $\left(p/\lgc\right)^3\ll1$, are sufficient to produce an overall attraction (see below). \ra{A detailed calculation of these terms is found in Appendix~\ref{appA}.} Throughout the calculation, we assume that $p/\lgc<1$. Namely,  that the interaction is not strong, because of either screening or small patch-charge densities. In Ref.~\cite{SilbertPRL}, for example, where the screening was weak due to low salinity, one finds that $p/\lgc\approx4$ for the largest surface patches. Our results will then apply only for higher but reasonable salt concentrations, $n_{b}\ge10\,\rm{mM}$.

After obtaining the potential $\Psi$, the free energy is more conveniently calculated using the charging method, and the osmotic pressure is derived via the thermodynamic relation $\Pi=-A^{-1}\left(\partial F/\partial d\right)$.  The average osmotic pressure is obtained by averaging over the uniform distribution of the phase $\alpha$: $\left\langle \Pi\right\rangle _{\alpha} =(2\pi)^{-1}\int_{0}^{2\pi}\Pi\left(\alpha\right)\D\alpha$. This is an average over the different possible phases, determined by the experimental setup.  As the surface-charge densities are assumed to be quenched, all phases have the same probability, \ra{and no favorable configuration dominates the interaction. Furthermore, as the surfaces are large enough, all possible phases should be realized across the surfaces, making the quenched average suitable for describing the net interaction between them.} For brevity, the averaging notation $\left\langle ...\right\rangle _{\alpha}$ will be omitted hereafter. In addition, we define a dimensionless rescaled osmotic pressure $\tilde{\Pi}\equiv 2\pi\lb\lgc^2\Pi/\kbt$.

\ra{The calculation of the osmotic pressure can be repeated by retaining higher orders in the expansion of Eq.~(\ref{eq6}).  However, such higher-order terms lead to only quantitative changes that become negligible in the limits discussed below; see Appendix~\ref{appA}. The calculation can also be repeated for different surface charge distributions. However, we argue that the simple one-mode approximation, Eq.~(\ref{eq1}), represents qualitatively well the general case of overall neutral surfaces with a typical patch size. This is supported by a simple extension of the single mode modulation presented in Appendix~\ref{appB}.}

\section{Results}
\label{sec3}
\subsection{General}
\label{ssce3a}

\begin{figure*}[ht]
\centering
\begin{subfigure}[b]{0.45\textwidth}
\includegraphics[width=0.85\textwidth]{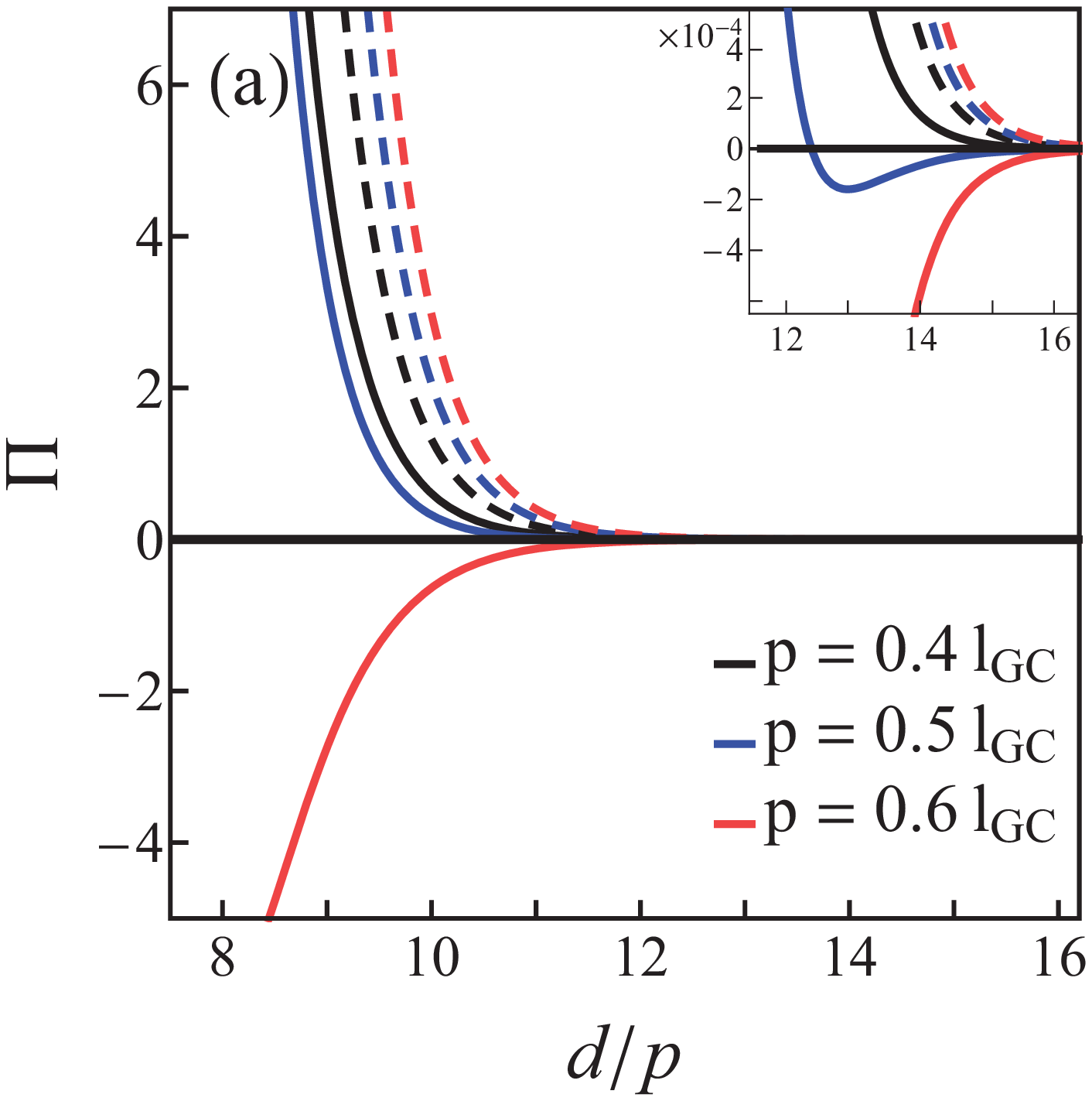}
\end{subfigure}
\begin{subfigure}[b]{0.45\textwidth}
\includegraphics[width=0.85\textwidth]{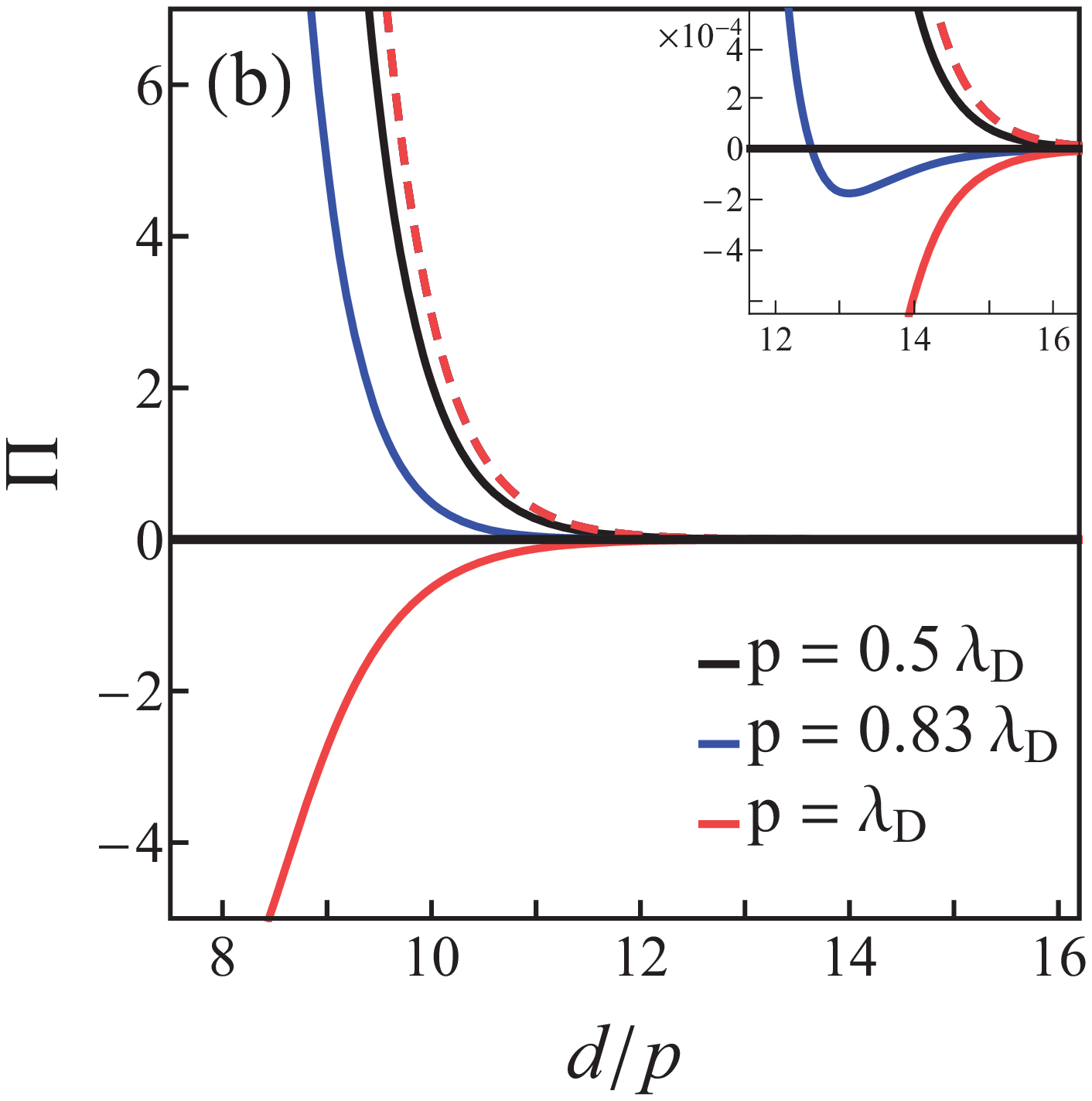}
\end{subfigure}
\caption{(Color online) Osmotic pressure profiles between patchy surfaces in units of $10^{-9}\times\kbt/(2\pi\lb p^{2})$. For $T=298\,\rm{K}$, $\eps=80$ and $p=1\,\rm{nm}$, this scaling factor corresponds to units of mPa.  Our calculation is plotted as solid curves, and the DH approximation is plotted as dashed ones. (a) Pressure profiles between surfaces with infinitely large charged patches ($p=\ld$) for three patch charge-densities, $p/\lgc=0.4,\,0.5$ and $0.6$. For $p=1\,\rm{nm}$, these three $p/\lgc$ values correspond to patch charge-densities $\sigma=e/11\,\rm{nm}^2$, $e/9\,\rm{nm}^2$ and $e/7\,\rm{nm}^2$.  (b) Pressure profiles between surfaces with a fixed patch charge-density ($p=0.6\,\lgc$) for three patch width values, $p/\ld=0.5,\,0.83$ and $1.0$. For fixed $p=1\,\rm{nm}$, the corresponding patch widths are $w=3.6\,\rm{nm}$, $5.6\,\rm{nm}$ and $w\to\infty$. In both (a) and (b), the intermediate profile (blue) crosses over from repulsion to attraction at smaller pressure values, as is shown in the corresponding insets. }
\label{fig2}
\end{figure*}

Results for the osmotic pressure are depicted in Fig.~\ref{fig2}. The dependence of the pressure on patch charge-density ($\sigma$), patch size ($w$), inter-surface separation ($d$) and bulk salt concentration ($n_{b}$) is expressed through the three ratios, $d/p$, $p/\lgc$ and $p/\ld$. Unlike the repulsive and monotonic pressure profiles of the DH theory (shown in the figure as dashed lines), in our calculation the pressure crosses over from repulsive to attractive at larger separations. Furthermore, the attraction can be  much stronger than the ever-existing van der Waals (vdW) attraction across the electrolyte, \ra{as is demonstrated in Sec.~\ref{ssec3c}}.

For small separations ($d/p\ll 1$), which yet satisfy $p/\lgc<d/p$, the DH term dominates, \ra{and we are left with the DH asymptotic form}
\begin{align}
\label{eq13}
\tilde{\Pi} &\approx \left(\frac{p}{d}\right)^2,
\end{align}
giving rise to a pressure that is purely repulsive and diverges in the $d\to0$ limit. This expression is derived for $p/\lgc<d/p$, because otherwise, higher order terms in Eq.~(\ref{eq6}) must be considered. The same requirement emerges in the DH limit for homogeneously and equally charged surfaces at small separations, which is valid only for $\ld/\lgc<d/\ld$ with $\tilde{\Pi}=4\left(\ld/d\right)^{2}$.
We see that in the patchy case, with zero net surface-charge, the repulsion is diminished, mostly due to the small factor $\left(p/\ld \right)^{2}$  that vanishes in the limit $p\sim w\to0$, corresponding to uncharged surfaces. We note that additional contributions were found from fluctuations beyond PB~\cite{Rudi06} for molecular-size patches. However, they still lead to an overall repulsion.

In the other limit of large separation ($d/p\gg1$), the osmotic pressure is found to be
\begin{align}
\label{eq14}
\tilde{\Pi} & \approx\left[4-\left(\frac{p^2}{\ld\lgc }\right)^{2}\frac{d}{p}\right]\e^{-2d/p}.
\end{align}
\ra{The first term in the right-hand side of Eq.~(\ref{eq14}) is the DH result, while the second term is the found correction.}
Two observations arise from Eq.~(\ref{eq14}).
The first is that the pressure becomes attractive at large separations, as is illustrated in Fig.~\ref{fig2}. Second, the exponential decay, $\exp(-2d/p)$, is faster as compared to the decay of $\exp(-d/\ld )$ for the homogeneous case. \ra{ The latter observation is evident already in the DH limit (for example, see  Refs.~\cite{Ohshima} and \cite{Richmond74})}, and has two origins:
(a)~the oppositely charged patches on each surface contribute to the screening,
as is evident from the replacement $\ld \to p$. (b)~The quenched average over the inter-surface phase, $\alpha$, eliminates a term of order $\exp(-d/p)$, leaving a leading term of order $\exp(-2d/p)$. \ra{This is the reason why the corrections to the DH result are important even in the limit of weak interaction. Once the term of order $\exp(-d/p)$ is averaged out, the previously negligible correction becomes substantial, and dominates for large separations.}

Equations~(\ref{eq13})~and~(\ref{eq14}) highlight clearly the qualitative similarities and differences between the heterogeneous and homogeneous surface-charge cases. Overall, the dependence on separation can be understood intuitively: at very small separations, the patches can be regarded as infinitely large. Accordingly, the repulsive osmotic pressure resembles that of homogeneously charged surfaces. At intermediate separations, the surface heterogeneity causes a  reduction of the free energy, which can lead to an overall attraction. As the separation is increased further, the free energy reduction decreases. Finally, at very large separations, the patchiness is smeared out, and the system can barely be distinguished from that of two uncharged surfaces.

\subsection{Repulsion-attraction crossover}
\label{ssce3b}

\begin{figure*}[ht]
\centering
\begin{subfigure}[b]{0.45\textwidth}
\includegraphics[scale=0.3]{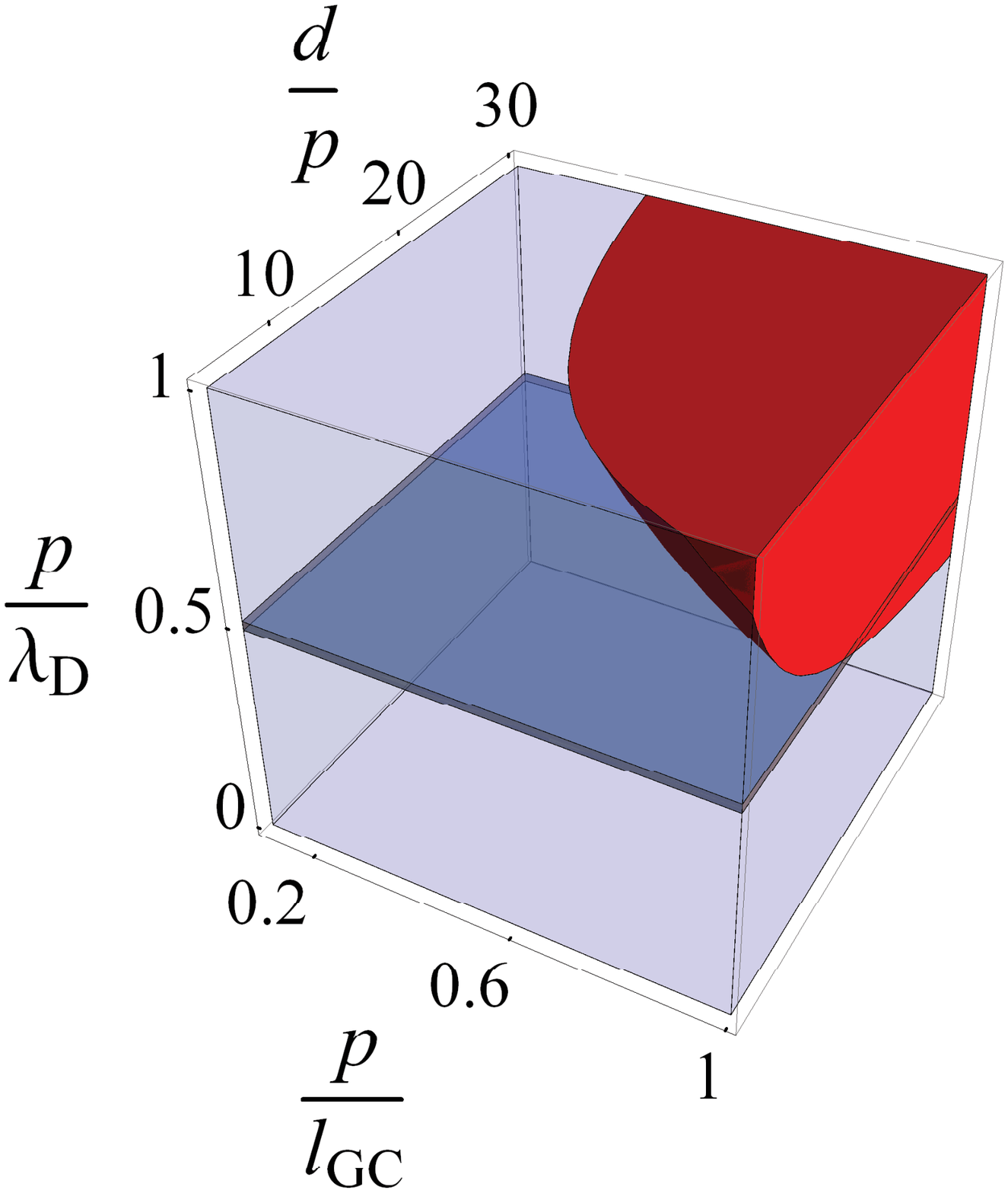}
\caption{}
\end{subfigure}
\begin{subfigure}[b]{0.2\textwidth}
\includegraphics[scale=0.3]{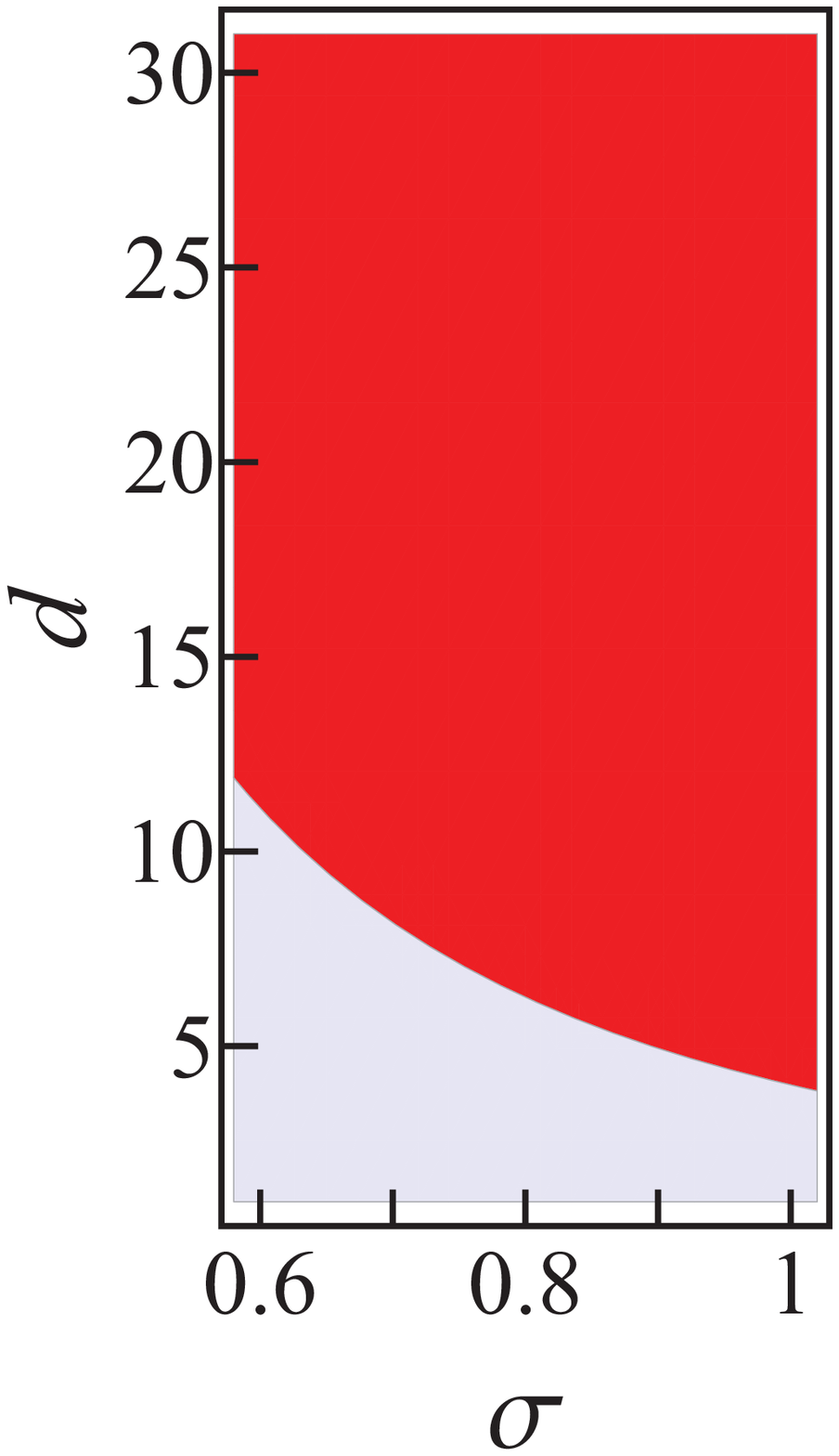}
\caption{}
\end{subfigure}\quad\quad
\begin{subfigure}[b]{0.2\textwidth}
\includegraphics[scale=0.3]{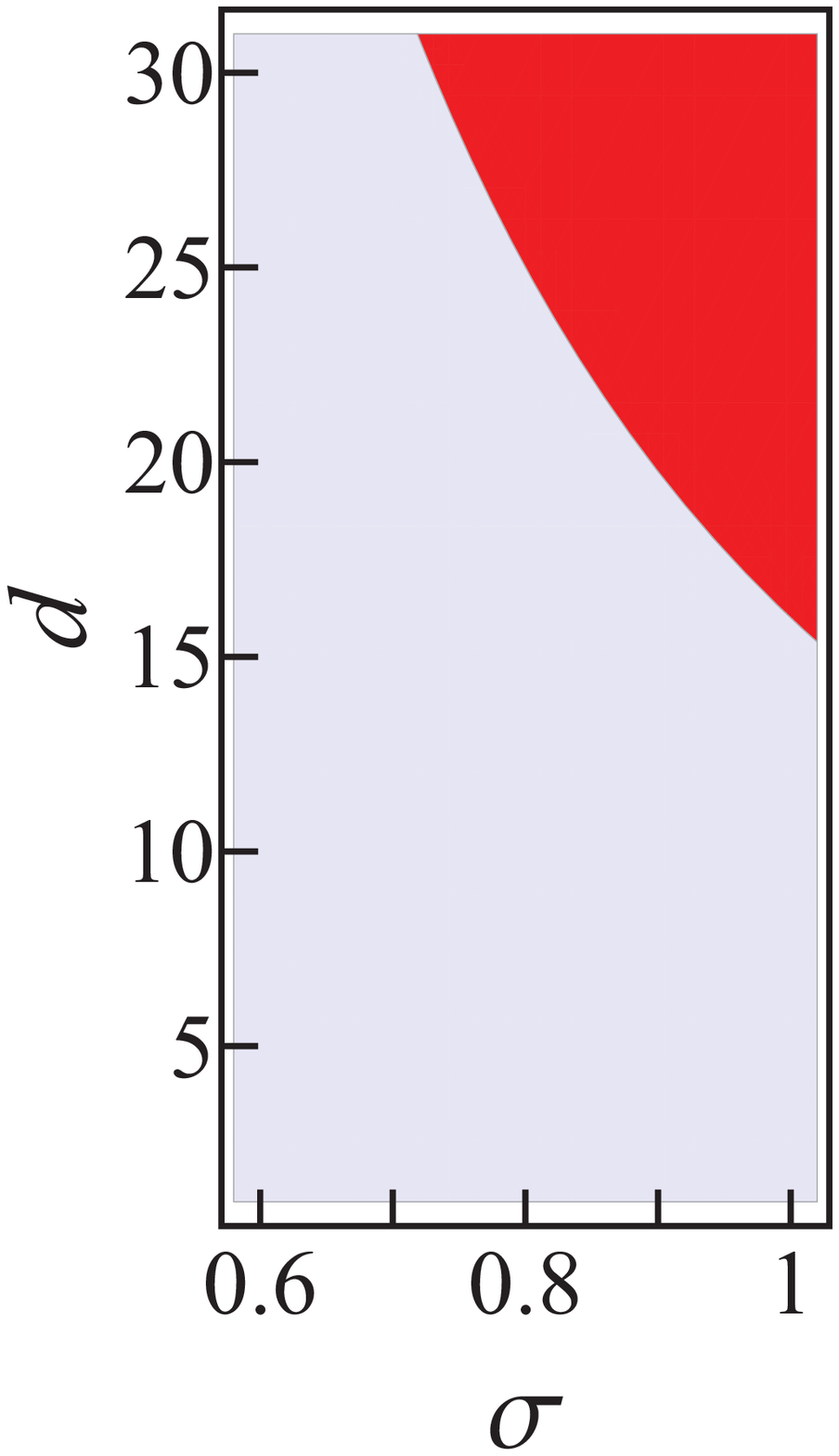}
\caption{}
\end{subfigure}
\caption{(Color online) Repulsion (blue or light) and attraction (red or dark) inter-surface interaction. (a) 3D plot in terms of the three ratios,  $p/\lgc$, $p/\ld$ and $d/p$.  The crossover occurs at smaller $d/p$ values with increased $p/\ld$ and increased $p/\lgc$. The ratio $p/\ld$ is large for large patches, while $p/\lgc$ is large for a combination of large patch charge-densities and large screening lengths. The two cuts in (b) and (c) are taken along the $p/\ld$ axis of (a), defining two planes of fixed $p/\ld$ values. (b) 2D cut for infinitely large charged patches ($p=\ld$ at the top face of (a)) plotted in terms of $\sigma$ in units of $e/(2\pi\lb\ld)$ and $d$ in units of $\ld$. (c) 2D cut for charged patches of a finite width with $p=\ld/2$ (mid-plane of (a) marked in dark gray), plotted in terms of $\sigma$ in units of $e/(\pi\lb\ld)$ and $d$ in units of $\ld/2$. The figure clearly demonstrates that the inter-surface attraction domain increases with the patch size, $w$. }
\label{fig3}
\end{figure*}

From Eq.~(\ref{eq14}) we find a crossover separation distance, $d^*$, between repulsion and attraction:
\begin{equation}\label{eq15}
d^{\ast}=4p^{-3}\left(\ld\lgc\right)^{2}=4[1+(\pi\ld /w)^{2}]^{3/2}\lgc ^{2}\ld^{-1}.
\end{equation}
This crossover is one of our major results and cannot be found within the DH framework. Note that the crossover separation indeed satisfies $d^{\ast}\gg p$ within our framework, justifying the use of Eq.~(\ref{eq14}) for its derivation. Equation~(\ref{eq15}) shows that for a fixed salt concentration, the crossover $d^{\ast}$ decreases with increased patch width, $w$, and increased patch charge-density, $\sigma$. In particular, the separation $d^*$ is minimal in the limit of infinitely large patches, where the electrostatic interaction is strong. Then, the crossover occurs at $d^{\ast\ast}\equiv d^{\ast}(w\to\infty)=4\lgc^{2}/\ld$. In the opposite limit of infinitesimally
small patches, the pressure becomes small and strictly positive as $d^{\ast}$ diverges.
For a physically accessible choice of parameters, such as $n_{b}=0.1\,\mbox{M}$ and $\sigma=e/5\,\mbox{nm}^{2}$ at $T=300\,K$, the crossover occurs at $d^{\ast}\simeq 6.2\,\mbox{nm}$ for a patch width of $w=10\,\mbox{nm}$,
and at $d^{\ast}\simeq 5.5\,\mbox{nm}$ for a patch width of $w=100\,\mbox{nm}$.

The crossover can be equally expressed in terms of the patch charge-density and patch width, which also separate the repulsion and attraction regions, as is illustrated in Fig.~\ref{fig3}. For example, beyond the minimal separation for attraction, {\it i.e.}  $d>d^{\ast\ast}$, the crossover patch width is
\begin{align}
\label{eq16}
w^{\ast}=\frac{\pi\ld}{\sqrt{\left(d/d^{\ast\ast}\right)^{2/3}-1}}.
\end{align}
Using the same choice of physical parameters as above, for a separation of $d=10\,\mbox{nm}$, the crossover from repulsion to attraction will occur for $w>w^{\ast}\simeq 4.2\,\mbox{nm}$, while
for a separation of $d=100\,\mbox{nm}$ it occurs for $w>w^{\ast} \simeq 1.2\,\mbox{nm}$.

\subsection{Comparison with the van der Waals attraction}
\label{ssec3c}
\ra{Aside the electrostatic attraction that we have dealt with, one should keep in mind the ever-present van der Waals (vdW) attraction between uncharged surfaces. While the first originates from the quenched averaged electrostatics between surface-charge patches, the latter stems from the thermal-averaged fluctuations of solvent dipoles.  As only the electrostatic attraction is sensitive to the surface-charge heterogeneity, the competition between the two is determined by varying the patch-charge density, $\sigma$, and patch width, $w$.}

\ra{In the presence of salt, the vdW attraction decays exponentially rather than algebraically~\cite{NinhamVDW,ParsegianVDW,NetzVDW}. In the limit of large separations, the vdW force per unit area, $f_{\rm{vdW}}$, is given by \cite{NinhamVDW}}
\be
\label{eq17}
\frac{f_{\rm{vdW}}}{\kbt}=-\frac{1}{4\pi\ld^3}\frac{\e^{-2d/\ld}}{d/\ld},
\ee
\ra{independent of $\sigma$ and $w$. It decays with a screening length of $\ld$, larger than the electrostatic screening length, $p$. However, the exponent in Eq.~(\ref{eq17}) is multiplied by a decreasing term $\sim d^{-1}$, as opposed to an increasing term $\sim d$ in Eq.~(\ref{eq14}). The interplay between these features will determine which of the two attractions is dominant.}

\ra{For large patches ($\ld\ll w$), the long-range electrostatic attraction is stronger than vdW for sufficiently large patch-charge densities and low salinity. Explicitly, comparing Eqs.~(\ref{eq14}) and (\ref{eq17}), we find that the electrostatic attraction is dominant for}
\be
\label{eq19}
\left(\frac{\pi\lgc}{w}\right)^{4}\le2\frac{\ld}{\lb} u^{2}\e^{-u},
\ee
\ra{where $u\equiv d\ld/ w^{2}$. The ratio on the left-hand side of Eq.~(\ref{eq19}) is inversely proportional to the total patch charge, while the ratio on the right-hand side depends solely on bulk properties and increases with the salt concentration.}

 \ra{As the function $f(u)=u^{2}\e^{-u}$ is bounded from above by about $0.5$, Eq.~(\ref{eq19}) implies that the long-range electrostatics are comparable with vdW only for $\left(\pi\lgc/w\right)^{4}<\ld/\lb$. Inserting $\lb/\ld<1/10$, as is realized in most experimental setups, this simplifies to $e/\left(|\sigma|w\right)<20\lb$. For reasonable physical values, the electrostatic term will then dominate over a finite range of $u$ values, corresponding to a large range of separations with $d=u w^{2}/\ld$. For example, for $T=300\,\rm{K}$, $n_{b}=2\,\rm{mM}$,  $e/\left(|\sigma|w\right)=3\,\rm{nm}$ and $w=100\,\rm{nm}$, the electrostatic term is dominant for separations up to $d=650\,\rm{nm}$. }

\section{Discussion and Conclusions}
\label{sec4}
In conclusion, we have found, within purely mean-field electrostatics (PB), that overall neutral patchy surfaces in solution {\em always} attract each other at sufficiently large separations. The attraction prevails not only for very large and strongly charged patches or low salinity, but also for a broad range of the system's controlled length scales, $d,\,w,\,\lgc$, and $\ld$.  Furthermore, for large patches, it is expected to be stronger than the vdW attraction. Our findings, therefore, suggest that the attraction effect plays a more important role than what has been perceived.

Our results reveal that the DH theory provides a qualitatively {\it wrong} picture of the interaction between overall neutral patchy surfaces with quenched surface charges. \ra{This is because the leading term in the DH result vanishes on average for the different possible patch displacements, rendering the initially small correction to the pressure significant at large separations.} We emphasize this point, as the calculation was performed in the limit of weak interactions, for which the linearized DH framework is usually justified. This limitation of the DH theory should be taken into account in the study of electrochemical systems that require an averaging over the screened electrostatic interaction.

In this paper we presented simple analytic expressions for the osmotic pressure between overall neutral surfaces with quenched charged surface patches. The results were derived at small and large inter-surface separations, as well as for charged patches of any size. These expressions should be useful in describing numerous physical systems where interacting charged surfaces are coated in patches by oppositely charged objects such as proteins, lipids and surfactants.

\vskip 0.5cm

{\it Acknowledgments.~}
We thank J. Dzubiella, J. Klein, P. Pincus, D. Pine, G. Silbert, and T. Witten for fruitful discussions, and T. Markovich and S. Safran for numerous suggestions. This work was supported in part by the Tel Aviv University -- Humboldt University Berlin
``{\it Biological and Soft Matter Physics}" joint cooperation program, the Israel Science Foundation (ISF) under Grant No. 438/12, the US-Israel Binational Science Foundation (BSF) under Grant No. 2012/060, and the ISF-NSFC joint research program under Grant No. 885/15. DA would like to thank the hospitality of the Free University (FUB), the Technical University (TUB) and the Humboldt University (HUB), Berlin, where this project was completed. He also thanks the Alexander von Humboldt Foundation for a research award.

\appendix
\section{The expansion in powers of $p/\lgc$}
\label{appA}
We base our results on an expansion in powers of $p/\lgc$ according to Eq.~(\ref{eq6}), where the odd powers in the potential, Eq.~(\ref{eq6}), lead to even powers in the free energy, Eq.~(\ref{eq3}). For each value of the inter-surface phase, $\alpha$ [Eq.~(\ref{eq1})], we find the osmotic pressure up to $4^{\rm{th}}$ order, and finally average over $\alpha$. Here, we present the detailed calculation of the expansion terms. The expansion is shown to converge, and  the next order is negligible in our discussed limits.

The expansion of Eq.~(\ref{eq6}) transforms the Euler-Lagrange equations of Eq.~(\ref{eq6a}) into the following  set of decoupled linear ordinary differential equations:
\begin{align}
\label{eq7}
h_{1}'' -h_{1} & = -2\cos\left(\frac{\alpha}{2}\right)\left[\zeta_{1}\delta\left(\tilde{z}+\tilde{d}/2\right)+
\zeta_{2}\delta\left(\tilde{z}-\tilde{d}/2\right)\right], \nonumber \\
 g_{1}'' -g_{1} & = -2\sin\left(\frac{\alpha}{2}\right)\left[\zeta_{1}\delta\left(\tilde{z}+\tilde{d}/2\right)-\zeta_{2}\delta\left(\tilde{z}-\tilde{d}/2\right)\right], \nonumber \\
 h_{3}'' -h_{3} & = \frac{1}{8}\left(\frac{p}{\ld}\right)^{2}\left(h_{1}^{2}+g_{1}^{2}\right)h_{1},
 \nonumber \\
 g_{3}'' -g_{3} & = \frac{1}{8}\left(\frac{p}{\ld}\right)^{2}\left(g_{1}^{2}+h_{1}^{2}\right)g_{1},
 \end{align}
where the argument of the $h$ and $g$ functions is $\tilde{z}\equiv z/p$ and $\tilde{d}\equiv d/p$. The 1st-order terms reproduce the Debye-Hu\"ckel solution with
\begin{align}
 \label{eq8}
 h_1(\tilde{z})&=2\cos\left(\frac{\alpha}{2}\right)\frac{\zeta_{1}\cosh\left(\tilde{z}-\tilde{d}/2\right)+\zeta_{2}\cosh\left(\tilde{z}+\tilde{d}/2\right)}{\sinh(\tilde{d})}
 \nonumber \\[3pt]
 & \nonumber\\
 g_1(\tilde{z})&=2\sin\left(\frac{\alpha}{2}\right)\frac{\zeta_{1}\cosh\left(\tilde{z}-\tilde{d}/2\right)-\zeta_{2}\cosh\left(\tilde{z}+\tilde{d}/2\right)}{\sinh(\tilde{d})}.
   \end{align}
Given $h_{1}$ and $g_{1}$, it is possible to evaluate $h_{3}$ and $g_{3}$  via the Green's function for the differential operator $\partial_{z}^{2}-1$ with the boundary condition of a vanishing electric field,
 \begin{align}
 \label{eqA2}
 G(z_1,z_2)=&-\frac{\Theta(z_1-z_2)}{\sinh(d)}\cosh(z_1-d/2)\cosh(z_2+d/2)\nonumber
 \\ &-\frac{\Theta(z_2-z_1)}{\sinh(d)}\cosh(z_2-d/2)\cosh(z_1+d/2),
 \end{align}
where $\Theta(z)$ is the Heaviside function.

The expressions for $h_{3}$ and $g_{3}$ are lengthy. For brevity, we present their values only at the boundaries. These expressions suffice to determine the free energy, due to the charging method [Eq.~(\ref{eq2})].
%
\begin{align}
 \label{eq9}
 h_{3}\left(-\tilde{d}/2\right) &=-\frac{\left(p/\ld\right)^2}{32\sinh(\tilde{d})^4}
 \left[\cos\left(\halpha\right)a\left(\tilde{d},\ZO,\ZT\right) \right.\nonumber\\
 &+ \left. \cos\left(\frac{3\alpha}{2}\right)b\left(\tilde{d},\ZO,\ZT\right)\right],\nonumber\\
 g_{3}\left(-\tilde{d}/2\right) &=-\frac{\left(p/\ld\right)^2}{32\sinh(\tilde{d})^4}
 \left[\sin\left(\halpha\right)a\left(\tilde{d},\ZO,-\ZT\right)\right. \nonumber\\
 &+ \left. \sin\left(\frac{3\alpha}{2}\right)b\left(\tilde{d},\ZO,-\ZT\right)\right],
 \end{align}
where
  \begin{align}
  \label{eq10}
a\left(d,\ZO,\ZT\right)\ = \ &
  \ZO^{3}\sinh(4d)+4d\ZO\left(3\ZO^{2}+4\ZT^{2}\right)\nonumber\\
+\ &4\ZO\sinh(2d)\left(2\ZO^{2}+5\ZT^{2}\right)\nonumber\\
+\ &11\ZT\sinh(d)\left(2\ZO^{2}+\ZT^{2}\right)\nonumber \\
+\ &3\ZT\sinh(3d)\left(2\ZO^{2}+\ZT^{2}\right)\nonumber\\
+\ &12d\ZT\cosh(d)\left(2\ZO^{2}+\ZT^{2}\right)\nonumber \\
+\ &8d\ZO\ZT^{2}\cosh(2d),\nonumber \\\nonumber\\
b\left(d,\ZO,\ZT\right) \ =\ &\ZO\ZT\Big[4\cosh(d)\left(3d\ZO+5\ZT\sinh(d)\right)\nonumber \\
+\ &2\ZO\sinh(d)(3\cosh(2d)+7)\nonumber \\
+\ &4d\ZT(\cosh(2d)+2)\Big].
  \end{align}
Similarly, the expressions for the terms evaluated at the top surface are obtained from Eq.~(\ref{eq9}) via the mapping $\ZO\leftrightarrow\ZT$.

The next order terms in Eq.~(\ref{eq6}) can be calculated in a similar manner. Consider the term $\left(p/\lgc\right)^{5}\left(\cos(kx)h_5(\tilde{z})+\sin(kx)g_5(\tilde{z})\right)$. Inserting these terms in the free energy, Eq.~(\ref{eq2}), and minimizing with respect to $h_{5}$ and $g_{5}$ yields the following ODEs:
\begin{align}
\label{eqA1}
h_{5}''-h_{5}=&\frac{1}{8}\left(\frac{p}{\ld}\right)^{2}\left(3h_{1}^{2}+g_{1}^{2}\right)h_{3}\nonumber
\\+&\frac{1}{192}\left(\frac{p}{\ld}\right)^{2}\left(h_{1}^{4}+g_{1}^{4}+2h_{1}^{2}g_{1}^{2}\right)h_{1},\nonumber
\\[3pt] g_{5}''-g_{5}=&\frac{1}{8}\left(\frac{p}{\ld}\right)^{2}\left(3g_{1}^{2}+h_{1}^{2}\right)g_{3}\nonumber
\\+&\frac{1}{192}\left(\frac{p}{\ld}\right)^{2}\left(g_{1}^{4}+h_{1}^{4}+2g_{1}^{2}h_{1}^{2}\right)g_{1}.
\end{align}
Once again, $h_{5}$ and $g_{5}$ are found by  using the Green's function, Eq.~(\ref{eqA2}).

\begin{figure}[ht]
\centering
\includegraphics[width=0.85\columnwidth]{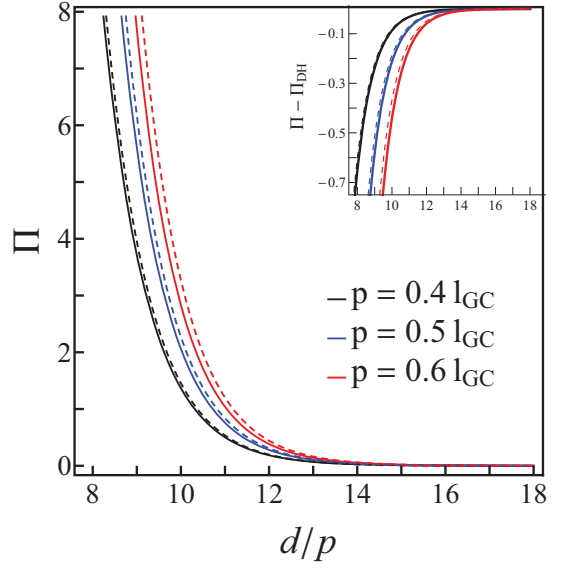}
\caption{(Color online) Osmotic pressure profiles plotted for in-phase surfaces ($\alpha=0$) and infinitely large patches ($w\to\infty$) in units of $10^{-5}\times\kbt/(2\pi\lb p^{2})$. Different patch-charge densities are denoted by different colors. The $4^{\rm{th}}$ order calculation is plotted as solid curves, and the DH approximation is plotted as dashed curves. The $6^{\rm{th}}$ order calculation coincides with the solid curves of the $4^{\rm{th}}$ order. In the inset, the deviation from the DH result is plotted as solid curves for the $4^{\rm{th}}$ order and as dashed curves for the $6^{\rm{th}}$ order one. Clearly, the expansion converges.}
\label{fig4}
\end{figure}

As the expressions for the functions $h_{5}$ and $g_{5}$ are lengthy, they are not presented here. Instead, in Fig.~\ref{fig4} we present osmotic pressure curves for in-phase surfaces ($\alpha=0$) for different orders of the expansion. It is evident that while the $4^{\rm{th}}$ order expansion differs from the $2^{\rm{nd}}$  order one (DH), it nearly coincides with the $6^{\rm{th}}$ order expansion, demonstrating convergence. Similar curves are obtained for different values of $\alpha$.

Moreover, for the average pressure, we show that the $6^{\rm{th}}$ order is indeed negligible in the appropriate limits. For small separations ($\tilde{d}\ll1$), the DH term dominates and Eq.~(\ref{eq13}) remains unchanged. For large separations ($\tilde{d}\gg1$), we find that
\begin{equation}
\label{eqA3}
\Pi\approx\left[4-\tilde{d}\left(\frac{p^2}{\ld\lgc }\right)^{2}\left(1-\frac{3}{4}\left(\frac{p^2}{\ld\lgc }\right)^{2}\right)\right]\e^{-2\tilde{d}}.
\end{equation}
Evidently, for sufficiently small $p^2/(\ld\lgc)$, the correction to Eq.~(\ref{eq14}) is negligible.

\section{The single mode approximation}
\label{appB}
The surface-charge densities in our model are described by  a single-mode modulation [Eq.~(\ref{eq1})].  This is a special case of the more general surface-charge density that can be described as a sum over Fourier modes,
\begin{align}
\label{eqB1}
\sigma_{1}(x) & =\sum_{n=1}^{N}\sigma_{n}\cos(k_{n}x-\alpha_{n}/2)\nonumber \\
\sigma_{2}(x) & =\sum_{n=1}^{N}\sigma_{n}\cos(k_{n}x+\alpha_{n}/2).
\end{align}
where $\{\sigma_n\}$ are taken as the {\it same} Fourier amplitudes of the bottom ($\sigma_1$) and top ($\sigma_2$) charged surfaces, and $\{\alpha_n\}$ are the relative phase shifts between the two.
In view of the general Fourier sum, our model, Eq.~(\ref{eq1}), can be considered as the limit where one mode dominates over the rest. We will now show that this limit provides a good approximation under some reasonable assumptions.

Consider surface-charge densities consisting of two modes, $k$ and $q$, with the same phase-shift $\alpha$,
\begin{align}
\label{eqB2}
\sigma_{1}(x)  &=\sigma_{k}\cos(kx-\alpha/2)+\sigma_{q}\cos(qx-\alpha/2),\nonumber\\
&\nonumber\\
\sigma_{2}(x)  &=\sigma_{k}\cos(kx+\alpha/2)+\sigma_{q}\cos(qx+\alpha/2).
\end{align}
We will assume that the $k$ mode is the dominant one, being the smaller mode, $k/q<1$, and having a larger surface charge-density amplitude, i.e.  $|\xi|=|\sigma_{q}/\sigma_{k}|<1$. As done in Sec.~\ref{sec2}, we insert a potential of the form, $\Psi=\Psi_{k}+\Psi_{q}$, in the free energy, Eq.~(\ref{eq2}), with
\begin{align}
\label{eqB3}
\Psi_{k}\left(x,z\right) & =\frac{p_{k}}{\lgc}\Bigl[\cos(kx)h\left(z/p_{k}\right)+\sin(kx)g\left(z/p_{k}\right)\Bigr],
&\nonumber\\ \nonumber\\
\Psi_{q}\left(x,z\right) & =\frac{p_{q}}{\lgc}\Bigl[\cos(qx)s\left(z/p_{q}\right)+\sin(qx)r\left(z/p_{q}\right)\Bigr],
\end{align}
where $p_{k}=1/\sqrt{\ld^{-2}+k^{2}}$ and $p_{q}=1/\sqrt{\ld^{-2}+q^{2}}$.

We minimize the free energy up to 3rd order in $\lgc^{-1}$ with respect to $h$, $g$, $s$ and $r$. Expanding the functions according to
\begin{align}
\label{eqB4}
\Psi_{k}\left(x,z\right) & =\cos\left(kx\right)\sum_{n\ \rm{odd}}\left(\frac{p_{k}}{\lgc }\right)^{n}h_{n}\left(z/p_{k}\right) \nonumber\\
&+\sin\left(kx\right)\sum_{n\ \rm{odd}}\left(\frac{p_{k}}{\lgc }\right)^{n}g_{n}\left(z/p_{k}\right),
\end{align}
 (and similarly for $\Psi_{q}$) and equating powers of $\lgc^{-1}$ results in a set of ordinary differential equations. For example, in terms of the argument $\tilde{z}\equiv z/p_{k}$ and $\tilde{d}\equiv d/p_{k}$, we find for the $k$ mode that
\begin{align}
 \label{eqB5}
 h_{1}''-h_{1}& = -2\cos\left(\frac{\alpha}{2}\right)\left[\zeta_{1}\delta(\tilde{z}+\tilde{d}/2)+\zeta_{2}\delta(\tilde{z}-\tilde{d}/2)\right], \nonumber \\
 g_{1}''-g_{1}& = -2\sin\left(\frac{\alpha}{2}\right)
 \left[\zeta_{1}\delta(\tilde{z}+\tilde{d}/2)-\zeta_{2}\delta(\tilde{z}-\tilde{d}/2)\right], \nonumber \\
 h_{3}-h_{3}& = \frac{1}{8}h_{1}\left(\tilde{z}\right)\left(\frac{p_{k}}{\ld}\right)^{2}\left[h_{1}^{2}\left(\tilde{z}\right)+g_{1}^{2}\left(\tilde{z}\right)\right] \nonumber \\
 &+\frac{1}{8}h_{1}\left(\tilde{z}\right)\left(\frac{p_{q}}{\ld}\right)^{2}
 \left[s_{1}^{2}\left(\frac{p_{k}}{p_{q}}\tilde{z}\right)+r_{1}^{2}
 \left(\frac{p_{k}}{p_{q}}\tilde{z}\right)\right],
 \nonumber \\
 g_{3}''-g_{3}& = \frac{1}{8}g_{1}\left(\tilde{z}\right)\left(\frac{p_{k}}{\ld}\right)^{2}\left[g_{1}^{2}\left(\tilde{z}\right)+h_{1}^{2}\left(\tilde{z}\right)\right] \nonumber \\
 &+\frac{1}{8}g_{1}\left(\tilde{z}\right)\left(\frac{p_{q}}{\ld}\right)^{2}
 \left[r_{1}^{2}\left(\frac{p_{k}}{p_{q}}\tilde{z}\right)+s_{1}^{2}
 \left(\frac{p_{k}}{p_{q}}\tilde{z}\right)\right].
 \end{align}
The equations for the $q$ terms are obtained via the transformation $k\to q$, $h_{n}\to s_{n}$, $g_{n}\to r_{n}$ and $\zeta_{i}\to\xi\zeta_{i}$. In particular, for the $q$ terms, the corresponding rescaled variables are $\tilde{z}\equiv z/p_{q}$ and $\tilde{d}\equiv d/p_{q}$.

Comparing Eq.~(\ref{eqB5}) with the one-mode equations [Eq.~(\ref{eq7})], we find that the equations (and consequently, their solutions) preserve their form, except a new inhomogeneous term that couples between the two modes (the second lines for the $h_3$ and $g_3$ expressions in Eq.~(\ref{eqB5})). In addition, from the charging method, Eq.~(\ref{eq3}), and using the fact the two modes are orthogonal, the free energy can be written as a sum $F=F_{k}+F_{q}$, where
\begin{align}
  \label{eqB6}
  F_{k} & =\int \D x\,\D y\,\int_{0}^{1}\psi_{k}\left(-\frac{d}{2};\ZO\sigma_{k},0\right)\sigma_{k}\cos\left(kx-\frac{\alpha}{2}\right)\D\ZO\nonumber \\
 & +\int \D x\,\D y\,\int_{0}^{1}\psi_{k}\left(\frac{d}{2};\sigma_{k},\ZT\sigma_{k}\right)\sigma_{k}\cos\left(kx+\frac{\alpha}{2}\right)\D\ZT
\end{align}
%
(and similarly for $F_{q}$), with $\psi_{k}(\pm d/2;\sigma,\sigma')$ being   the $k$-mode electrostatic potential at the $z=\pm d/2$ surfaces, given that the bottom (top) surface-charge density amplitude is $\sigma$ ($\sigma'$).
Consequently, the total osmotic pressure can be written in the form $\Pi=\Pi_{k}+\Pi_{q}+\Pi_{kq}$, where $\Pi_{k}$ and $\Pi_{q}$ are obtained from solving the corresponding one-mode surfaces, and $\Pi_{kq}$ originates from the new inhomogeneous terms.

\begin{figure*}[ht]
\centering
\begin{subfigure}[b]{0.45\textwidth}
\includegraphics[width=0.9\textwidth]{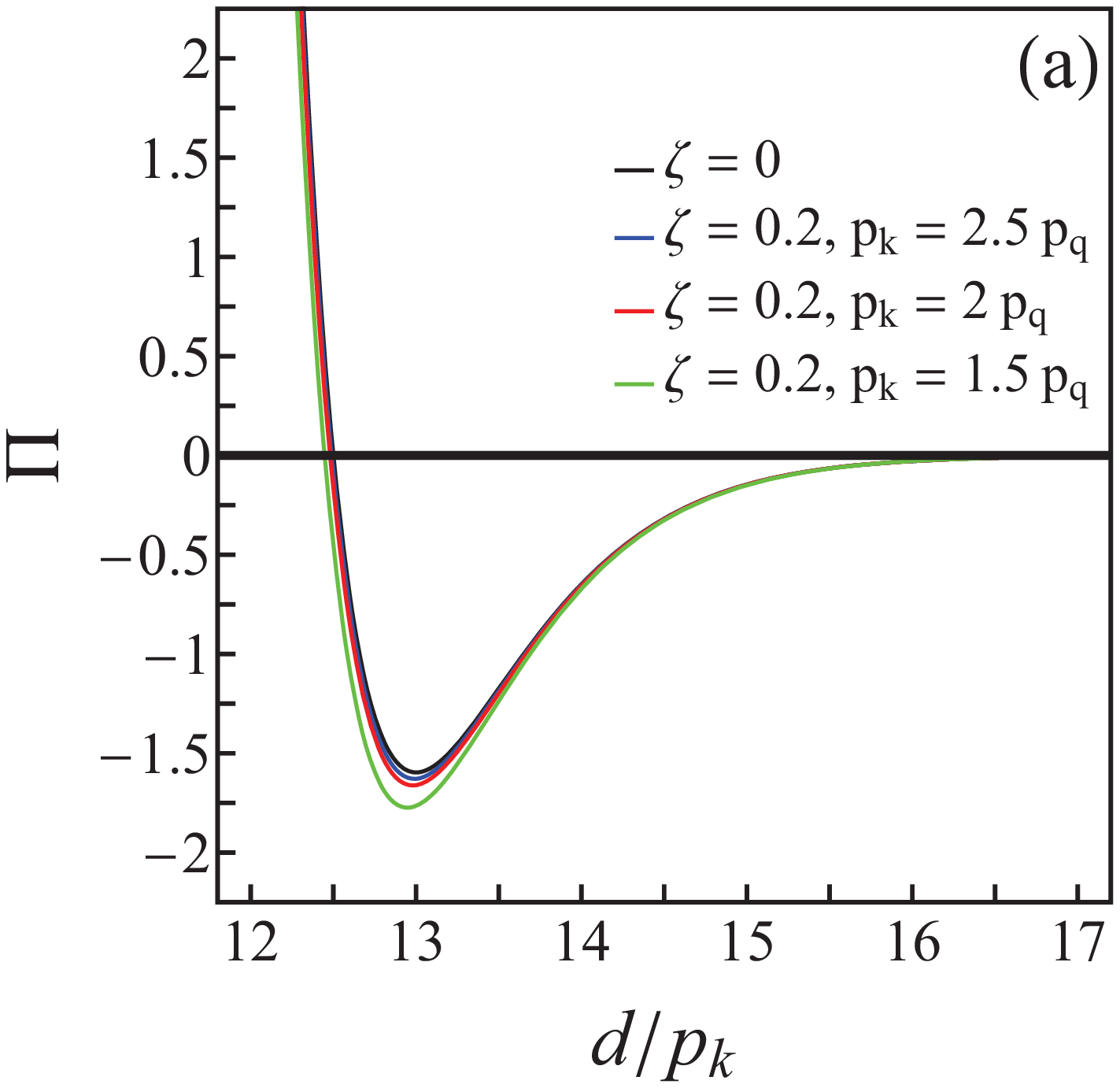}
\end{subfigure}
\quad
\begin{subfigure}[b]{0.45\textwidth}
\includegraphics[width=0.9\textwidth]{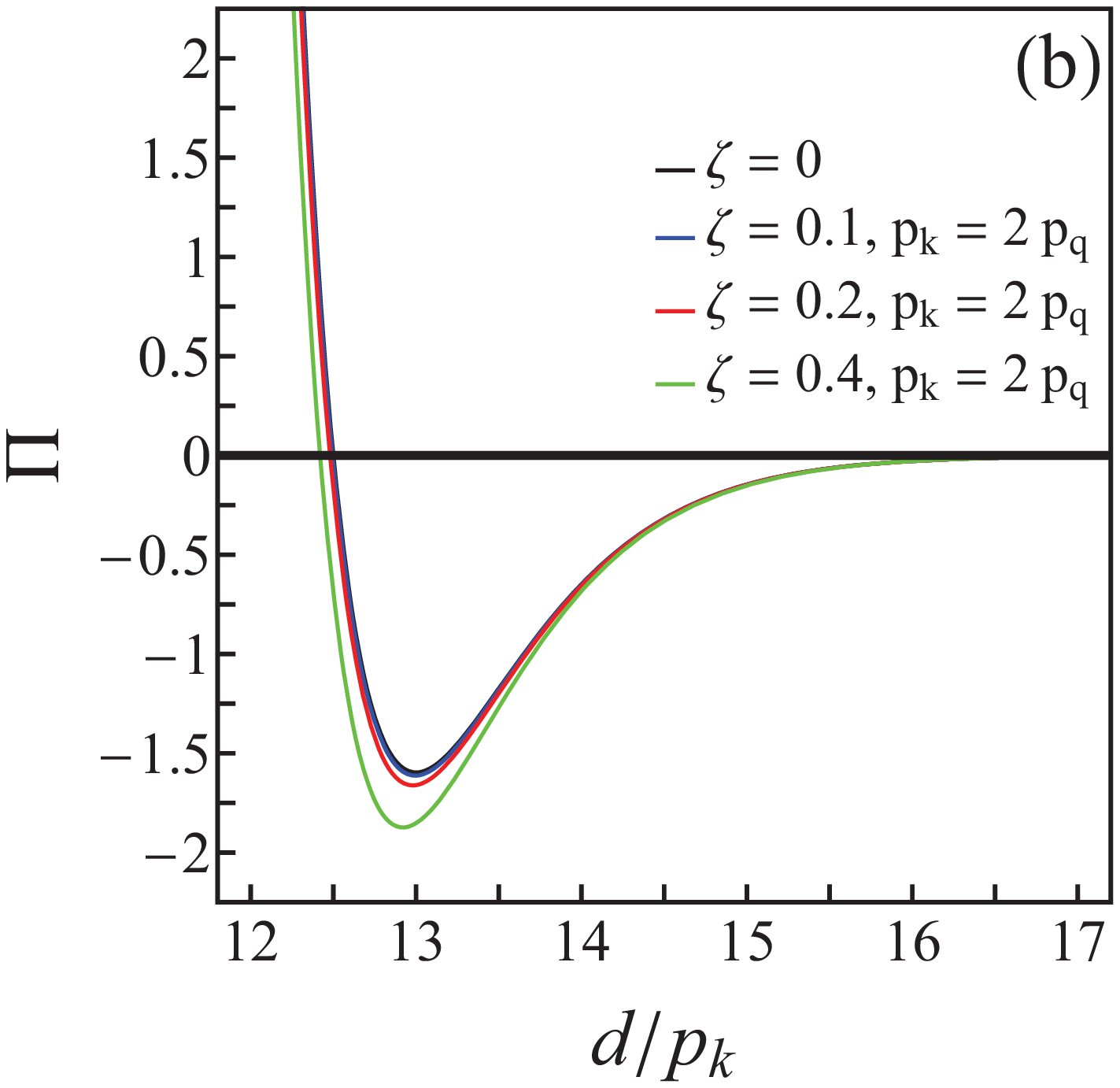}
\end{subfigure}
\caption{(Color online) Osmotic pressure profiles plotted for two-mode ($k$ and $q$) surface-charge densities, as compared to the one-mode approximation ($\xi=0$) in units of $10^{-13}\times\kbt/(2\pi\lb p^{2})$. (a) Different wavenumbers $k$ and $q$, and fixed surface charge density parameter, $\xi$. (b) Different surface charge density parameter. $\xi$, and fixed wavenumbers $k$ and $q$. It is evident that the one-mode result is a good approximation for small $\xi$ values and large $p_{k}/p_{q}$ values.}
\label{fig5}
\end{figure*}

It is clear from our results for the osmotic pressure, Eqs.~(\ref{eq13})~and~(\ref{eq14}), that the term $\Pi_{q}$ is subdominant for $|\xi|<1$ and $p_{k}/p_{q}>1$. Evidently, the same holds for $\Pi_{kq}$ that can only lead to a stronger long-range attraction at large separations in these limits. This is illustrated in Fig.~\ref{fig5} for different values of $\xi$ and $p_{k}/p_{q}$. Therefore, the one-model is a good approximation for surface charge densities with a dominant mode, or equivalently, systems with a dominant patch size.

\newpage

\end{document}